\documentclass[12pt,psfig]{article}
\usepackage{amssymb}
\usepackage{amsmath,amsthm}
\usepackage{amsfonts}
\usepackage{graphicx}
\usepackage{braket}
\usepackage{graphics}
\usepackage{indentfirst}
\usepackage{hyperref}

\begin{document}

%%%%%%%%%%%%%%%%%%%%%%%%%%%%%%%%%%%
%%%%%%%%%%%%%%%%%%%%%%%%%%%%%%%%%%%
%%%%%%%%%%%%%    THE TITLE PAGE    %%%%%%%%%%%
%%%%%%%%%%%%%%%%%%%%%%%%%%%%%%%%%%%
%%%%%%%%%%%%%%%%%%%%%%%%%%%%%%%%%%%

    \begin{center}
        {\Large\textbf{Thermodynamics and Tachyon Condensation 
        of the Dressed-Dynamical Unstable D$p$-branes
        }}\\
\vspace*{.95cm}
{\Large  \textsc{Hamidreza Daniali}}\\

\vspace*{.75cm}
 {Department of Physics, Amirkabir University of
Technology (Tehran Polytechnic) \\
P.O.Box: 15875-4413, Tehran, Iran \\}
Email: \ \textit{hrdl@aut.ac.ir}
    \end{center}
\vspace*{1.25cm}
    \begin{abstract}
    Using the boundary state formalism and 
    thermo field dynamics approach,
    we study a D$p$-brane at finite temperature
    in which the background 
    Kalb-Ramond field, a $U(1)$ gauge potential,
    and tachyon field are turned on together
    with a general tangential dynamics.
    The thermal entropy of the brane will be studied.
    In addition, the behavior of the
    entropy after the tachyon condensation process
    will be investigated and
    some thermodynamic interpretations will be extracted.
    \end{abstract}
\begin{flushleft}
\vspace*{.25cm}
\end{flushleft}

\newpage

%%%%%%%%%%%%%%%%%%%%%%%%%%%%%%%%%%%%%
%%%%%%%%%%%%%%%%%%%%%%%%%%%%%%%%%%%%%
%%%%%%%%%%%%%%%%    THE BODY    %%%%%%%%%%%%%
%%%%%%%%%%%%%%%%%%%%%%%%%%%%%%%%%%%%%
%%%%%%%%%%%%%%%%%%%%%%%%%%%%%%%%%%%%%
\section{Introduction}
\label{100}
%%%%%%%%%%%%%%%%%%%%%%%%%%%%%%%%%%%%%
D$p$-branes are absolutely necessary to comprehend string theory
and its relationship to field theories and gravity \cite{1}.
Since the introduction of D$p$-branes, some of the most fundamental
physical results in string theory have been discovered by examining
their properties \cite{2, 3}.
In addition, they have significantly contributed to
our grasp of dualities \cite{1, 3}.
Furthermore, by adding dynamics, various backgrounds,
and internal fields to the brane, several
intriguing features of the D$p$-branes can be revealed
through so-called boundary state formalism
\cite{4}-\cite{24}.

Among various configurations, thermalizing
D$p$-branes have been a considerable focus of research.
On the one hand, the relationship between
D$p$-branes and field theory at finite
temperature is an intriguing subject in
and of itself that might aid in our
comprehension of the physical features of D$p$-branes.
In the low energy limit of string theory,
where D$p$-branes are solitonic
solutions to supergravity,
some investigation has been made in this area.
In this limit, the thermodynamics of
D$p$-branes have been stated within
the context of path-integral field
theory formulation at finite temperature
\cite{25}-\cite {30}. On the other hand, they have been
utilized to understand the statistical features
of different systems, such as the Hawking temperature,
energy-entropy relation and Hagedorn transition of extreme,
near-extreme, and Schwarzschild black holes
\cite{31}-\cite{43}.
Despite the relative understanding of strings at finite temperature
and the promising discoveries from D$p$-brane ensembles,
little is known about the statistical features of D$p$-branes.

To implement temperature to the structure of D$p$-branes,
one must be able to modify, at the Fock space level,
an adaption from zero temperature to finite temperature.
Since in the boundary state language, the D$p$-brane is expressed
in terms of string operators acting
on the vacuum, a convenient way is to employ the
Thermo Field Dynamics (TFD) formalism \cite{44}, \cite{45}.
Uniting such an approach with the D-branes boundary state
formalism and generalizing it to include more extending
configurations has been accomplished in remarkable
Refs. \cite{46}-\cite{51}.
Previously, TFD was used to examine
the renormalization of open bosonic
strings at finite temperature.
Also, the compatibility of the
renormalization with the thermal Veneziano
amplitude has been demonstrated \cite{52}.
In Ref. \cite{53}, the global
phase structure of the thermal bosonic
string ensemble and its connection to the
thermal string amplitude are explained.

In addition to thermalizing D$p$-branes,
the existence of an open string tachyon on it
inherently renders it unstable,
which is another fascinating topic in string theory.
Sen conjectured that the open string
tachyon condensation represents the decay of unstable
D$p$-branes into the closed string vacuum \cite{54}.
In other words, Through the process of tachyon condensation,
An unstable D$p$-brane subsequently either decays to a stable
D$(p-1)$-brane or collapses to the closed string vacuum.
This indicates that D$p$-branes may be established as
a source of closed string \cite{55}-\cite{57}.
Using the boundary state and tachyon
condensation, it is possible to determine the
time evolution of the source for
each mode of a closed string \cite{58}.
Also, it has been claimed that the boundary
state description of the rolling tachyon
is valid for a finite duration dictated by string coupling,
after which we may encounter energy
dissipation into the bulk \cite{59}.
Such physical phenomena, which are
involved with the decay of an object, resulted
in the development of the background-independent
string theory formulation.

The thermal feature of the D$p$-brane,
conjointly with the turned-on tachyonic
field on it, motivated and stimulated
us to study the entropy of the brane.
However, to construct the most general thermal boundary state,
some other extensions have been also implemented.
Precisely, in this paper, we calculate
the entropy of the bosonic D$p$-brane dressed with a $U(1)$
 gauge potential and a tachyonic profile
 in the presence of the background $B$-field
 which is thermalized in the context of the TFD approach.
 We also considered general tangential dynamics to the brane.
 In addition, the entropy of the brane
 is also computed after tachyon condensation. As will be shown,
 since the tachyon condensation leads to a 
 change in value of the entropy,
 the second law of thermodynamics
 will also be examined for our system.

This paper is organized as follows.
In Sec. \ref{200}, the boundary state corresponding
to a dressed-dynamical unstable D$p$-branes
at zero temperature is reviewed.
In Sec. \ref{300}, we shall construct the
general thermal boundary state
when all fields in our configuration are turned on together with
general tangential dynamics. We calculate the entropy of the
D$p$-brane. In Sec. \ref{400}, at first, the
tachyon condensation is briefly introduced, and,
after that, the effect of the tachyon
condensation on the entropy is studied.
Besides, some thermodynamical interpretations will be provided.
Sec. \ref{500} is devoted to the conclusions.

%%%%%%%%%%%%%%%%%%%%%%%%%%%%%%%%%%%%%%
\section{Review of the boundary state}
\label{200}
%%%%%%%%%%%%%%%%%%%%%%%%%%%%%%%%%%%%%%

In this section, the boundary state corresponds to an 
unstable-dressed D$p$-brane with tangential
dynamics in the zero temperature, $T=0$, is introduced.
Consequently, we start with the following
sigma-model action for closed string
\begin{eqnarray}
\label{1}
S =&-&\frac{1}{4\pi\alpha'} {\int}_\Sigma
{\rm d}^{2}\sigma \left(\sqrt{-\mathcal G}\mathcal G^{ab}
G_{\mu\nu} +\varepsilon^{ab} B_{\mu\nu}\right)
\partial_a X^{\mu}\partial_b
X^{\nu}
\nonumber\\
&+&\frac{1}{2\pi\alpha'} {\int}_{\partial\Sigma}
{\rm d}\sigma \left( A_\alpha
\partial_{\sigma}X^{\alpha}+
\omega_{\alpha\beta}J^{\alpha\beta}_{\tau }
+T(X^\alpha)\right),
\end{eqnarray}
where $\Sigma$ is the closed string worldsheet,
while $\partial\Sigma$ indicates its boundary.
To avoid dealing with ghosts, we choose
to work in the light-cone gauge $X^0 \pm X^{d-1}$. Hence,
$\mu,\nu \in \{1 , \cdots , d-2\}$, $\alpha,
\beta \in \{1 , \cdots ,p \}$
and $i, j \in \{p+1 , \cdots ,d-2 \}$ are spacetime indices,
worldvolume directions of the D$p$-brane,
and its perpendicular directions in
light-cone coordinates, respectively.
Additionally, the metrics of the target spacetime are denoted by
$\eta_{\mu\nu}= {\rm diag}(-1,+1,\cdots,+1) $,
whereas $\mathcal G_{ab}$ with $a,b\in\{\tau,\sigma \}$
signifies the metrics of the string worldsheet.
The background fields are the constant Kalb-Ramond field
$B_{\mu\nu}$, the $U(1)$ internal gauge field $A_\alpha$
and the open string tachyon field $T(X^\alpha)$.
In order to preserve the quadratic structure of
the action to become path integrally solvable,
we employ the Landau gauge
$A_{\alpha}=-\frac{1}{2}F_{\alpha \beta }X^{\beta}$
with the constant field strength
$F_{\alpha \beta }$, and the tachyon profile
$T= \frac{1}{2}U_{\alpha\beta}X^{\alpha}X^{\beta}$
where $U_{\alpha\beta}$ is a constant symmetric matrix.
The constant antisymmetric tensor
$\omega_{\alpha \beta}$ with the explicit form
${\omega }_{\alpha \beta}J^{\alpha\beta }_{\tau }=
2{\omega }_{\alpha \beta}X^{\alpha}{\partial }_{\tau
}X^{\beta}$
designates the angular velocity which expresses the tangential
dynamics of the brane.
Due to the presence of background fields on the worldvolume
of the brane, the Lorentz symmetry has been manifestly broken.
Thus, the tangential dynamics of the brane in the directions of
its worldvolume is sensible.

By setting the variation of the action to zero
we receive the equation of motion and
the following boundary state equations
\begin{eqnarray}
\label{2}
&~& \left[(\eta_{\alpha \beta} + 4\omega_{\alpha \beta})
{\partial}_{\tau }X^{\beta}
+{\mathcal F}_{\alpha \beta}
{\partial }_{\sigma }X^{\beta}
+U_{\alpha \beta }X^{\beta }\right]_{\tau =0}\ \
|\mathcal B\rangle\ =0,
\nonumber\\
&~& {\delta X}^i|_{\tau =0}|\mathcal B\rangle\ =0,
\end{eqnarray}
where the total field strength is
${\cal{F}}_{\alpha \beta}=F_{\alpha \beta}-B_{\alpha \beta}$.

The well-known closed string mode expansion
simply allows us to rewrite
Eqs. \eqref{2}, in terms of the zero modes and oscillators.
The coherent state method enables us to obtain the solution
of the oscillatory portion of the boundary state equations
\begin{eqnarray}
\label{3}
{|\mathcal B\rangle}^{({\rm osc})}\ =\prod^{\infty }_{n=1}
{[\det Q_{(n)}]^{-1}}{\exp \left[-\sum^{\infty }_{m=1}
A^{\dagger\mu }_{m}\Omega_{(m)\mu \nu }
B^{\dagger\nu }_{m}\right]
{|0\rangle}_{\alpha }
\otimes{|0\rangle}_{\beta }}\;,
\end{eqnarray}
where we have used
$\alpha^\mu_n$ and $\beta^\mu_n$ as
left- and right-moving oscillators, respectively.
Furthermore, the subsequent notations were used
\begin{eqnarray}
A^\mu_n = \frac{\alpha_n^\mu}{\sqrt{n}}, &\qquad& A^{\dagger\mu}_n
= \frac{\alpha_{-n}^\mu}{\sqrt{n}}, \nonumber\\
B^\mu_n = \frac{\beta_n^\mu}{\sqrt{n}}, &\qquad& B^{\dagger\mu}_n
= \frac{\beta_{-n}^\mu}{\sqrt{n}},\label{4}
\end{eqnarray}
for $n>0$ with the algebras
\begin{eqnarray}
&& [A^\mu_n , A^{\dagger\nu}_m] =  [B^\mu_n , B^{\dagger\nu}_m]
= \delta_{n,m} \eta^{\mu\nu}, \nonumber \\
&& [B^\mu_n , A^{\dagger\nu}_m] =  [A^\mu_n , B^{\dagger\nu}_m]
= {\rm all \ other\  commutators} = 0.\label{5}
\end{eqnarray}
The normalization factor
$\prod^{\infty }_{n=1}{{[\det Q_{(n)}]}^{-1}}$ in Eq. \eqref{3}
comes from the disk partition function \cite{21}.
The mode-dependent matrices possess the following definitions
\begin{eqnarray}
&~& Q_{(m){\alpha \beta }} =
\eta_{\alpha \beta} + 4\omega_{\alpha \beta}-
{{\mathcal F}}_{{\mathbf \alpha }{\mathbf \beta
}}+\frac{i}{2m}U_{\alpha \beta },\label{6}\\
&~& \Omega_{(m)\mu\nu}=(\Delta_{(m)\alpha \beta}\;
,\; -{\delta}_{ij}),\label{7}\\
&~& \Delta_{(m)\alpha \beta} =
(Q_{(m)}^{-1}N_{(m)})_{\alpha \beta}, \label{8}\\
&~& N_{(m){\alpha \beta }} = \eta_{\alpha \beta}
+ 4\omega_{\alpha \beta}
+{{\mathcal F}}_{{\mathbf \alpha }{\mathbf \beta }}
-\frac{i}{2m}U_{\alpha \beta }.
\label{9}
\end{eqnarray}
For the zero-mode part of the boundary state,
by using higher dimensional Gaussian integral,
we receive the following expression
\begin{eqnarray}
\label{10}
{{\rm |}\mathcal B\rangle}^{\left(0\right)}&=&
\sqrt{\dfrac{(2\pi)^p}{\det{\mathcal{R}}}}
\prod^p_{\alpha=1} {\rm |}p^{\alpha }\rangle
\prod_i{\delta {\rm (}x^i}{\rm -}y^i{\rm )}
{\rm |}p^i\rangle\;,
\end{eqnarray}
where $\mathcal R_{\alpha\beta}$ possesses the definition
\begin{eqnarray}
&~& \mathcal R_{\alpha \beta }=-2i \alpha'\big[{\mathcal U}+
U^{-1}(\eta +4\omega) +{(\eta+4\omega)}^T
U^{-1}\big]_{\alpha \beta },
\nonumber\\
&~& {\mathcal U}{\mathbf =}\left( \begin{array}{ccc}
{\left(U^{-1}{(\eta +4\omega)}\right)}_{11} & \cdots
& {\mathbf 0}\\
\vdots  & \ddots  & \vdots  \\
{\mathbf 0} & \cdots  & {\left(U^{-1}{(\eta +4\omega)}
\right)}_{pp}
\end{array} \right).
\nonumber
\end{eqnarray}
In the light-cone gauge,
one can construct the total boundary state as
\begin{eqnarray}\label{11}
{|\mathcal B\rangle}^{(\rm tot)}=\frac{\mathcal T_p}{2}
{|\mathcal B\rangle}^{(\rm osc)}\otimes
{|\mathcal B\rangle}^{(0)},
\end{eqnarray}
where $\mathcal T_p$ is the D$p$-brane tension.
%%%%%%%%%%%%%%%%%%%%%%%%%%%%%%%%%%%%%
\section{The entropy}
\label{300}

In this section, at first, the boundary states
corresponding to a dressed-dynamical
unstable D$p$-branes at finite temperatures are constructed.
Given that the Eqs. \eqref{2}-\eqref{11} are
represented in terms of bosonic string operators and states,
these entities must be merely mapped at $T \ne 0$.
An intriguing way to do that is by
utilizing TFD approach.
To ensure that the content of this paper
may be understood without being
sidetracked by excessive computations,
all fields of the brane are independent of temperature.

According to the TFD approach, the thermodynamics of the
system is given in an extended Fock space consisting of the
original Fock space and an identical copy of it
(this is indeed true for the path-integral approach).
The total thermic system is comprised of the
original string and its duplicate indicated as `tilde' strings.
These two copies are independent and the total Fock space is
$\mathcal H_{\rm total} = \mathcal H \otimes \tilde{\mathcal{H}}$.
To execute this construction in the context of
the bosonic string theory,
we use $\tilde{A}^\mu_n, \tilde{B}^\mu_n$ and
... as identical operators
belonging to the copy of the system with independent algebras.

From now on we use the usual notation in
the literature to compute the entropy.
According to the Refs. \cite{44}-\cite{48},
$| \ \rangle\rangle$ signifies a
state from $\mathcal H_{\rm total}$.
The vacuum state is given by
$|0\rangle\rangle^{\rm (osc)} \equiv
|0\rangle\rangle^{\rm (osc)}_\alpha \otimes
|0\rangle\rangle^{\rm (osc)}_\beta$.
Owing to the fact that two Fock spaces are independent,
we may write vacuums as
$|0\rangle\rangle^{\rm (osc)}_{(\alpha,\beta)}
= |0\rangle^{\rm (osc)}_{(\alpha,\beta)}
\otimes \widetilde{|0\rangle}^{\rm (osc)}_{(\alpha,\beta)}$.
One should note that to have a total vacuum state,
the enlarging procedure must be also applied to the zero-modes.

Thermal features of the system in
$\mathcal H$ space are accomplished by introducing a set
of Bogoliubov unitary operators, i.e,
\begin{eqnarray}\label{12}
|0(\beta_T)\rangle\rangle_{(\alpha, \beta)} = \prod_{n>0}
e^{-i \mathcal G_n^{(\alpha,\beta)}}
|0\rangle\rangle_{(\alpha, \beta)},
\end{eqnarray}
in which
\begin{eqnarray}
\mathcal{G}_n^\alpha = -i \theta({\beta_T})
\left( A_n .\tilde  A_n - A^\dagger_n
.\tilde A^\dagger_n\right),\label{13}\\
\mathcal{G}_n^\beta = -i \theta({\beta_T})
\left( B_n .\tilde B_n - B^\dagger_n
.\tilde B^\dagger_n\right),
\label{14}
\end{eqnarray}
acting on the states and on the operators of the enlarged space.
In Eqs. \eqref{13} and \eqref{14}, $\beta_T = (k_BT)^{-1}$ where
$k_B$ the Boltzmann's constant and $ \theta({\beta_T})$ is a
temperature parameter whose value depends on the mode statistics.
Since we merely deal with the bosonic string theory,
the value of  $ \theta({\beta_T})$  is
\begin{eqnarray}
\cosh \theta_n  ({\beta_T}) &=& u_n ({\beta_T})
= \frac{1}{\sqrt{1-e^{-\beta_T{w_n}}}}, \nonumber\\
 \sinh \theta_n  ({\beta_T}) &=& v_n ({\beta_T})
 = \sqrt{\frac{e^{-\beta_T{w_n}}}{1-e^{-\beta_T{w_n}}}}. \label{15}
\end{eqnarray}
Given the fact that the Bogoliubov
operators do not combine the left- and right-moving states,
it also is possible to produce a direct product of the states as
$|0(\beta_T)\rangle\rangle = |0(\beta_T)\rangle\rangle_\alpha \otimes
|0(\beta_T)\rangle\rangle_\beta$.
The action of Bogoliubov transformations on oscillator operators
translates them to new temperature-dependent
operators through the Heisenberg picture,
\begin{eqnarray}
\{A_n^\mu(\beta_T), \tilde A_n^\mu(\beta_T),
B_n^\mu(\beta_T), \tilde B_n^\mu(\beta_T)\} =
e^{-i \mathcal G_n^\alpha} \{A_n^\mu, \tilde
A_n^\mu, B_n^\mu, \tilde B_n^\mu\}
e^{i \mathcal G_n^\alpha}, \label{16}
\end{eqnarray}
in which the following results can be received
\begin{eqnarray}
A_n^\mu(\beta_T) &=& u_n(\beta_T) A_n^\mu - v_n(\beta_T)
\tilde A_n^{\dagger \mu}, \qquad
\tilde A_n^\mu(\beta_T) = u_n(\beta_T) \tilde A_n^\mu
- v_n(\beta_T) A_n^{\dagger \mu},\label{17} \\
B_n^\mu(\beta_T) &=& u_n(\beta_T) B_n^\mu - v_n(\beta_T)
\tilde B_n^{\dagger \mu}, \qquad
\tilde B_n^\mu(\beta_T) = u_n(\beta_T) \tilde B_n^\mu
- v_n(\beta_T) B_n^{\dagger \mu}.\label{18}
\end{eqnarray}

At finite temperature, there are three
possible formulations for D$p$-brane.
One of them, which follows in this paper,
is to map all associated operators
and states to their thermal counterparts
(for other possibilities and relations between them,
see Ref. \cite{47}).
Therefore, the boundary state of dressed
dynamical unstable D$p$-branes at $T\ne 0$ is
\begin{eqnarray}
&&| \mathcal B (\beta_T)\rangle\rangle^{\rm (tot)} = \dfrac{\mathcal
T_p^2 }{4}  \dfrac{(2\pi)^{p}}{\det{\mathcal{R}}}
 \left\{\prod^{\infty }_{n=1} {[\det Q_{(n)}]^{-2}}\right\}
 \prod^p_{\alpha=1} {\rm |}p^{\alpha }\rangle
 \widetilde{{\rm |}p^{\alpha }\rangle}
\prod_{i= p+1}^{d-2}{\delta {\rm (}x^i}{\rm -}y^i{\rm )}
\widetilde{{\delta {\rm (}x^i}{\rm -}y^i{\rm )}}
{\rm |}p^i\rangle \widetilde{{\rm |}p^i\rangle}\nonumber \\
&~&\qquad {\exp \left[-\sum^{\infty }_{m=1}
A^{\dagger }_{m}(\beta_T).\Omega_{(m) }
.B^{\dagger }_{m}(\beta_T)\right]} {\exp \left[-\sum^{\infty }_{k=1}
\tilde A^{\dagger }_{k}(\beta_T).\Omega_{(k) }
.\tilde B^{\dagger }_{k}(\beta_T)\right]}
{|0(\beta_T)\rangle}.\quad
\label{19}
\end{eqnarray}

Now we study thermal properties of this system.
This can be done by introducing the entropy operator in the
context of TFD and sandwiching it between the established
boundary states \eqref{19}, i.e,
\begin{eqnarray}
\label{20}
\mathcal S = k_B \ ^{\rm(tot)}\langle\langle
\mathcal B (\beta_T)|
\mathcal K( A , B ; \theta (\beta_T)
| \mathcal B (\beta_T))\rangle\rangle^{\rm (tot)} .
\end{eqnarray}
Since we are dealing with a closed string,
the operator $\mathcal K$ takes the feature
\begin{eqnarray}
\mathcal K( A, B ; \theta (\beta_T)) &=&
\sum_{n=1}^\infty \Big[ (A_n^\dagger . A_n
+ B_n^\dagger . B_n) \ln \sinh^2 \theta_n (\beta_T)\nonumber \\
&-& (A_n. A_n^\dagger + B_n . B_n^\dagger)
\ln \cosh^2 \theta_n(\beta_T)\Big].\label{21}
\end{eqnarray}
Eqs. \eqref{20} and \eqref{21} can be
rewritten to include tilde strings.
However, given the postulation of Refs. \cite{44}, \cite{45},
to achieve the physical properties of the system,
the computation of  $\tilde{\mathcal S} = k_B \
^{\rm(tot)}\langle\langle \mathcal B (\beta_T)|
\tilde{\mathcal K}( A , B ; \theta (\beta_T)
| \mathcal B (\beta_T))\rangle\rangle^{\rm (tot)} $
must be dropped out.

By plugging Eqs. \eqref{19} and \eqref{21} into
Eq. \eqref{20}, The entropy of the dressed-dynamical
unstable D$p$-brane is
\begin{eqnarray}
\mathcal S &=& \mathcal T_p^2  k_B \frac{ (d-2)\mathcal
V_{d-2} \tilde{\mathcal V}_{d-2}}{ (2\pi)^{2(d-2p-2)}}
{\det}^{-1}_{p\times p}{(\mathcal R^\dagger \mathcal R)}
\left[\prod^\infty_{s=1} [{\rm det}_{p\times p}
Q_{(s)}]^{-4}\right] \nonumber \\
&\times & \sum_{n=1}^\infty \xi_n(\theta_n(\beta_T))
\prod_{m=1}^\infty\Big[{\det}\Big(
\mathbb{I} - \Omega_{(m)}^\dagger \Omega_{(m)}\Big)\Big]^{-2},
\label{22}
\end{eqnarray}
where $\mathcal V_{d-2}$ and $\tilde{\mathcal V}_{d-2}$
are the volume of the $\mathcal H$-space and
$\tilde{\mathcal H}$-space in light-cone gauge, respectively, and
\begin{eqnarray}
	\xi_n(\theta_n(\beta_T)) \equiv  \ln \sinh^2 \theta_n(\beta_T) 
	+ (1+3 \sinh^2 \theta_n(\beta_T)) \ln \tanh^2 \theta_n(\beta_T),
\end{eqnarray}
is the thermal function associated with the entropy. 
In the $T \rightarrow 0$ limit,
the contribution of the oscillators diverges.
In contrast, in $T \rightarrow \infty$ limit,
the oscillator's contribution is proportional to $\ln(-1)$.
This may indicate that the
concept of temperature breaks down at arbitrarily high
temperatures owing to similar phenomena that
happen at Hagedorn temperature in string theory.
Some plausible explanations are provided in Refs. \cite{46}, \cite{47}
in order to comprehend these values of entropy.

Another interesting result is that if we simply turn on the tachyon 
profile and off all background and internal fields, together with 
the tangential dynamics of the D$p$-brane, the resulted entropy will be zero. 
This may demonstrate the difference between 
the tachyonic instability and the thermal instability of the D-brane.

Due to the presence of the tachyon
field and the general tangential dynamics
of the branes, our results are more
general than those presented in Refs. \cite{46}, \cite{47}. 
However, by quenching the tachyonic field,
the matrix $\Delta_{(n)}$ becomes mode-independent,
and the results in mentioned reference are obtained when
$\omega_{\alpha\beta} \rightarrow 0$ is also applied.
With the attainment of entropy, 
it becomes feasible to deduce various 
other thermodynamic properties. 
Given the renormalization problem of the string tension from a 
finite-temperature renormalization-group approach, 
however, calculating the free energy, which helps us better 
comprehend phase transition and thermodynamic stability 
of the system, is quite difficult in the 
context of TFD \cite{52}, \cite{60}, \cite{61}. 
This may indicate future research that must be conducted 
within the context of D$p$-branes. 
%%%%%%%%%%%%%%%%%%%%%%%%%%%%%%%%%%%%%
\section{The effects of the tachyon condensation on the entropy}
\label{400}
%%%%%%%%%%%%%%%%%%%%%%%%%%%%%%%%%%%%%

When D$p$-branes are studied in the
presence of open string tachyonic fields,
instability of it, whose investigation is crucial to our
understanding of the vacuum of string theory, will take place.
Through so-called tachyon condensation,
a phase transition follows.
During this process, the D$p$-brane collapses drastically,
and we are left with a collection of closed strings.

From a mathematical standpoint, at
least one element of the tachyon
matrix $U_{\alpha\beta}$ must be
infinite to form the tachyon condensation.
For simplicity, we impose the condensation of the
tachyon field just in the $x^p$-direction,
that is, $U_{pp}\rightarrow \infty$.
There are three matrices involving
the tachyonic field in Eq. \eqref{22};
$\mathcal R$, $ Q_{(n)}$ and $\Delta_{(n)}$,
in which the limit must be considered.
By implementing $U_{pp} \rightarrow \infty$, we have
$\lim_{U_{pp} \rightarrow \infty} (U^{-1})_{p\alpha}=
\lim_{U_{pp} \rightarrow \infty} (U^{-1})_{\alpha p} = 0$.
Consequently, the matrix $\mathcal R$ loses its final row and column.
Let us denote it as $\bar{\mathcal{R}}$. 
Similarly, the effect of tachyon condensation on the
component $\prod^{\infty }_{n=1}
{[\det_{p\times p} Q_{(n)}]^{-4}}$, with the use of zeta
function regularization, becomes
\begin{eqnarray} \label{23}
\lim_{U_{pp} \rightarrow \infty}  \prod^{\infty }_{n=1}
{[{\rm det}_{p\times p}
Q_{(n)}]^{-4}} =  \pi^2 U_{pp}^2
\prod_{n=1}^\infty \Big[{\rm det}_{(p-1)\times 
(p-1)}Q^{[p-1]}_{(n)}\Big]^{-4},
\end{eqnarray}
where the matrix $Q^{[p-1]}_{(n)}$ can be obtained by removing the 
final row and column of matrix $Q_{(n)}$, 
resulting in a $(p-1)\times(p-1)$ matrix. 
The limit of the matrix $\Delta_{(n)}$,
since it is not the product of the limits of
$Q^{-1}_{(m)}$ and $N_{(m)} $, must be determined after
the executing the product $Q^{-1}_{(m)} N_{(m)} $. It gives rise to
\begin{eqnarray}
{\mathop{\lim }_{U_{pp}\to \infty}
\Delta_{(m)}}=\left(\begin{array}{cc}
\left(\mathbf\Delta_{(m)}\right)_{(p-1)\times (p-1)}
& $\;\;${\bf 0}_{(p-1)\times 1} \\
{\bf 0}_{1 \times (p-1)} & $\;\;$-1
\end{array} \right),
\end{eqnarray}
where 
\begin{eqnarray}
	\left(\mathbf\Delta_{(m)}\right)_{(p-1)\times (p-1)}
	= \left( Q^{[p-1]}_{(m)} \right)^{-1} N^{[p-1]}_{(m)}.
\end{eqnarray}
Now it is evident that a Neumann direction ($x^p$-direction)
has been altered into a Dirichlet direction.
Hence, the tachyon condensation phenomenon deforms
an unstable D$p$-brane into a stable D$(p-1)$-brane.

Adding all these together, the effect of the tachyon
condensation on the entropy, Eq. \eqref{22}, takes the feature
\begin{eqnarray}
\mathcal S' &=& (d-2)k_B  \frac{(U_{pp} \mathcal T_p)^2 \mathcal
V_{d-2} \tilde{\mathcal V}_{d-2}}{ 4(2\pi)^{2(d-2p-3)}}
{\det}^{-1}_{(p-1)\times (p-1)}{(\bar{\mathcal R}^\dagger 
\bar{\mathcal R})}
\left[\prod^\infty_{s=1} [{\rm det}_{p\times p}
Q^{[p-1]}_{(s)}]^{-4}\right] \nonumber \\
&\times & \sum_{n=1}^\infty \xi_n(\theta_n(\beta_T))
\prod_{m=1}^\infty\Big[{\det}\Big(
\mathbb{I} - \mathbf\Omega_{(m)}^\dagger 
\mathbf\Omega_{(m)}\Big)\Big]^{-2}
 ,\ \ \label{25}
\end{eqnarray}
where $ \mathbf \Omega_{(m)\mu\nu} \equiv \left\{\left[\left( 
Q^{[p-1]}_{(m)} \right)^{-1} N^{[p-1]}_{(m)}\right]_{\alpha'\beta'}\;
,\; -{\delta}_{i'j'}\right\}
$ in which $\alpha'
,\beta' \in \{1, \ldots, p-1\}$ and $i', j' \in\{p, \ldots d-2\} $.
From Eq. \eqref{25}, one can deduce that
portions that consist of the thermal factor remain the same.
However, due to the tachyon condensation, the
terms corresponding to the fields
and dynamics are drastically changed which
results in a new value for entropy.
Nevertheless, Eq. \eqref{25} is not just the entropy of a stable brane.
It comprises both the entropy of
D$(p-1)$-brane and the entropy associated
with the closed strings that were produced
during the collapse of the original brane.
As can be seen, entropy \eqref{25} is a
complicated function in terms of temperature,
fields and dynamics of the configuration,
making their separation rather impossible.

%%%%%%%%%%%%%%%%%%%%%%%%%%%%%%%%%%%%%
\subsection{The second law of the thermodynamics}
\label{401}
%%%%%%%%%%%%%%%%%%%%%%%%%%%%%%%%%%%%%

In this part, we explored some thermodynamic
aspects of the process of tachyon condensation.
Since the system evolves from its initial state,
i.e, the original D$p$-brane,
to its final state, which is the D$(p-1)$-brane
and the released closed strings,
studying the second law of thermodynamics must be interesting.
Hence, we must examine the validity of the inequality
$\Delta \mathcal S \equiv \mathcal S' - \mathcal S > 0$.
Given that in Eqs. \eqref{25} and \eqref{22}, the determinant factors
are always positive, in the limit of
$U_{pp} \rightarrow \infty$, the inequality is valid if and only if 
$\xi(\theta_n(\beta_T))$ is positive which signals the 
positivity of the entropy. 
This implies that
\begin{eqnarray}\label{27}
	\ln ( e^{\beta_T \omega_n} -1) 
	+ \beta_T \omega_n \frac{e^{\beta_T} 
	+ 2}{e^{\beta_T} -1} <0.
\end{eqnarray}
In the other words, the second law of 
thermodynamics holds true when 
the frequency value of the $n$-th oscillator, 
denoted as $\omega_n$, are within the range of 
$nk_B T \ln X$ where $X \approx (1, \sqrt{\pi/3}) $. 
This condition significantly limits the frequency 
levels of it.

It should be noted that, in the context of TFD approach, 
the attainment of a physically admissible configuration 
that adheres to the principles of the second law of thermodynamics 
necessitates a positive entropy. 
Consequently,  the instability of the brane, due to the 
presence of an open string tachyonic field on it, 
should not cause a negative value in entropy.
%%%%%%%%%%%%%%%%%%%%%%%%%%%%%%%%%%%%%
\section{Conclusions}
\label{500}

We presented a boundary state, corresponding
to a thermal-dynamical D$p$-brane
in the presence of an internal $U(1)$ gauge potential,
an open string tachyon
field and a non-zero constant Kalb-Ramond field.
For implementing the thermal aspect to the setup,
the temperature has been encoded
in the operators using the TFD approach.
This enabled us to map all operators in $T =0$ to $T\ne 0$.
Then, the thermal entropy of such branes was computed.
By assessing the resulting entropy at $T\rightarrow 0$, 
we observed that the entropy diverges.
Besides, in the limit $T \rightarrow \infty$,
the oscillator contribution
to the entropy is proportional to $\ln(-1)$.
Owing to the similar phenomena in the string theory at the
Hagedorn temperature, this may imply that the concept of
temperature breaks down at arbitrary
high temperatures.

Note that in simpler setups
similar conclusions have been also provided in the Refs.
\cite{46}-\cite{48}. This clarifies that, in the context
of the TFD approach, the resultant entropy
in the limits $T\rightarrow 0$ and
$T \rightarrow \infty$ is independent of
any field and dynamics.

We examined the effect of the
tachyon condensation in our setup and in the thermal entropy.
Since the thermal factors, associated with the entropy are unaffected
by the tachyon condensation, the study of the entropy
in the limits of low- and high-temperatures
yields the same results as the setup before
the tachyon condensation.
Despite this, as it can be seen in Eq. \eqref{25},
the effect of this phenomenon profoundly
manifests in the fields and dynamic contributors of entropy.

Finally, we examined the existence of the second law of 
thermodynamics for our setup during the tachyon condensation process. 
We observed that if D$p$-brane is in a thermal non-equilibrium state, 
the second law of thermodynamics 
will be violated after tachyon condensation. 
Therefore, to maintain a physical system, as long as one considers 
tachyon condensation, the D$p$-brane must not possess negentropy. 

%%%%%%%%%%%%%%%%%%%%%%%%%%%%%%%%%%%%%
\subsection*{Acknowledgement}
The author would like to thank D. Kamani for the
useful discussions and for reading the manuscript.

%%%%%%%%%%%%%%%%%%%%%%%%%%%%%%%%%%%%
%%%%%%%%%%%%%%%%%%%%%%%%%%%%%%%%%%%%
%%%%%%%%%%%%    THE BIBLIOGRAPHY    %%%%%%%%%%%
%%%%%%%%%%%%%%%%%%%%%%%%%%%%%%%%%%%%
%%%%%%%%%%%%%%%%%%%%%%%%%%%%%%%%%%%%

\end{document}